\RequirePackage{fix-cm}
\documentclass[smallextended]{svjour3}       % onecolumn (second format)
\smartqed  % flush right qed marks, e.g. at end of proof
\usepackage{graphicx}
\usepackage{amsmath,graphicx}
\usepackage{array}
\usepackage{multirow}
\usepackage{bm}
\begin{document}

\title{An End-to-End Approach to Automatic Speech Assessment for Cantonese-speaking People with Aphasia%\thanks{Grants or other notes
}


\author{Ying Qin         \and
        Yuzhong Wu \and
        Tan Lee \and
        Anthony Pak Hin Kong
        %etc.
}


\institute{Ying Qin \at
              Department of Electronic Engineering, The Chinese University of
Hong Kong, Shatin, N.T., Hong Kong \\
              %Tel.: +852-63532248\\
              \email{yingqin@link.cuhk.edu.hk}           %  \\
           \and
           Yuzhong Wu \at
              Department of Electronic Engineering, The Chinese University of
Hong Kong, Shatin, N.T., Hong Kong\\
             \email{yzwu@link.cuhk.edu.hk}
             \and
           Tan Lee \at
              Department of Electronic Engineering, The Chinese University of
Hong Kong, Shatin, N.T., Hong Kong\\
             \email{tanlee@cuhk.edu.hk}
             \and
           Anthony Pak Hin Kong \at
              Department of Communication Sciences and Disorders, University of Central Florida, Orlando, FL, USA\\
             \email{antkong@ucf.edu}
}

\date{Received: date / Accepted: date}

\maketitle

\begin{abstract}
Conventional automatic assessment of pathological speech usually follows two main steps: (1) extraction of pathology-specific features; (2) classification or regression on extracted features. Given the great variety of speech and language disorders, feature design is never a straightforward task, and yet it is most crucial to the performance of assessment. This paper presents an end-to-end approach to automatic speech assessment for Cantonese-speaking People With Aphasia (PWA). The assessment is formulated as a binary classification task to discriminate PWA with high scores of subjective assessment from those with low scores. The sequence-to-one Recurrent Neural Network with Gated Recurrent Unit (GRU-RNN) and Convolutional Neural Network (CNN) models are applied to realize the end-to-end mapping from fundamental speech features to the classification result. The pathology-specific features used for assessment can be learned implicitly by the neural network model. Class Activation Mapping (CAM) method is utilized to visualize how those features contribute to the assessment result. Our experimental results show that the end-to-end approach outperforms the conventional two-step approach in the classification task, and confirm that the CNN model is able to learn impairment-related features that are similar to human-designed features. The experimental results also suggest that CNN model performs better than sequence-to-one GRU-RNN model in this specific task.
\keywords{Pathological speech assessment \and End-to-end \and Aphasia \and Cantonese \and Deep neural network}
\end{abstract}
% ------------------------------------------------------------------------
% ------------------------------------------------------------------------
\section{Introduction}
\label{intro}
Aphasia refers to an acquired neurogenic speech-language disorder resulting from physical damage to specific brain regions. Symptoms of aphasia may adversely affect different modalities of language skills such as auditory comprehension, verbal expression, reading and writing \cite{benson1996aphasia}. The impairment could also span over various levels and components of the language system, including phonology, lexicon, syntax, and semantics \cite{Adam2014}. Speakers with aphasia may have difficulties in recalling names of objects and/or putting words together into sentences \cite{Aph_define}. Symptoms like anomia (word retrieval difficulty), dysfluency, voice disorder and dysprosody may be present in People With Aphasia (PWA) at various severity level and with different combinations \cite{wiki:xxx,wiki:xxxx}.
\par Speech assessment is an essential part of the comprehensive assessment for people with
aphasia (PWA), which aims at determining the type and/or severity degree of impairment. It is required to be carried out by a well-trained speech and language therapist, based on subjective evaluation of various abilities of language communication. Subjective assessment of speech from PWA is a challenging task because it requires not only clinical knowledge about the disease but also good understanding of relevant linguistic and cultural background. There are clearly urgent practical needs to develop effective and reliable methods of automatic speech assessment for PWA.
\par Automatic analysis of PWA speech is expected to facilitate objective and efficient assessment to assist diagnosis and rehabilitation of PWA. Automatic speech recognition (ASR) technique and machine learning based approaches have been applied in this area.
Various acoustic features and text features are designed and evaluated for PWA speech assessment. Peintner et al. \cite{peintner2008learning} performed an automatic classification on three sub-types of frontotemporal lobar degeneration in a relatively small dataset which includes the progressive non-fluent aphasia. They proposed a number of phone duration features, part-of-speech features as well as linguistic inquiry and word count features, which were extracted from ASR outputs. Fraser et al. \cite{fraser2013using,fraser2014automated} performed an automatic classification of sub-types of primary progressive aphasia using acoustic features and text features extracted from manual transcriptions. In their follow-up work \cite{fraser2013automatic}, an off-the-shelf commercial ASR system was adopted to automatically generate transcriptions for text feature extraction. However, the effectiveness of assessment was limited by the
low recognition accuracy on impaired speech. Duc Le et al. \cite{le2016improving,le2018automatic}  improved the PWA speech recognition using discriminative pre-training with out-of-domain dataset and multi-task acoustic
model. They achieved good performance in predicting subjective assessment
scores based on ASR outputs by analyzing text statistics (e.g. the number of nouns and verbs in spoken utterances), part-of-speech language model etc. Kohlschein et al. \cite{kohlschein2018automatic} proposed to automatically classify four most prevalent aphasia syndromes (Global aphasia, Broca's aphasia, Wernicke's aphasia and amnesic aphasia) based on manual transcriptions. A Long Short-Term Memory (LSTM) based classifier was trained with word vectors which were derived from the manual transcriptions. The peak classification accuracy obtained by their proposed model was $44.3\%$, and involving more training samples was suggested to further improve the classification performance.
The above studies target the automatic speech assessment for English-speaking and German-speaking PWA. In recent years, we made great efforts to develop an automatic assessment system for Cantonese-speaking PWA. In \cite{lee2013analysis}, the effectiveness of supra-segmental duration features in speech assessment of Cantonese-speaking PWA was investigated. The features such as duration of pause-delimited speech segments, duration of silence segments and frequency count of silence segments were found to be useful in differentiating speech of PWA from that of unimpaired individuals. The computation of these duration features requires a forced alignment process relying on manual transcriptions, which is undesirable in practical applications. In \cite{Qin2018icassp}, we proposed a framework of fully automatic speech assessment for PWA. Supra-segmental duration features were computed from time alignment generated by a dedicated ASR system. The text features were derived from the ASR-generated text output using syllable-level embedding technique. With the combination of these extracted features, the prediction of severity of impairment was formulated as a regression task. The prediction results showed a high correlation with subjective assessment scores.  We also attempted to improve the ASR performance on PWA speech using multi-task learning strategy, and used N-best lists and confusion networks to mitigate the effect of ASR errors on the text features \cite{Qin2018IS}.
\par From previous studies, it is evident that the exploration of pathology-specific features is an essential part of designing the assessment system and it is usually based on expert knowledge. However, some impairment-related characteristics contained in the raw speech data may be missed after the extraction of human-designed features. After extracting a great number of acoustic features and text features, post-processing procedures such as feature selection and classification/regression are required. They are generally complicated and time-consuming. In addition, parameters of feature extraction  and assessment model are optimized separately instead of jointly. Recently, the so-called
``end-to-end'' approach has demonstrated good successes in  ASR \cite{graves2014towards},  machine translation \cite{wu2016google}, and other applications. It requires relatively
little human effort on designing tailored features and shows superior performance \cite{graves2014towards}. Inspired by the ``end-to-end'' paradigm, we propose the design of a ``utterance-to-score'' system for PWA speech assessment to boost the efficiency of assessment system. Utterances spoken by PWA are fed directly to a Deep Neural Network (DNN), without requiring explicit feature extraction and selection. The output of the DNN gives the predicted assessment score for each input utterance. The overall assessment score for the impaired speaker is obtained by combining all his/her utterance-level scores.
\par This paper reports our preliminary results on PWA speech assessment using the end-to-end approach. Among numerous neural network structures, Recurrent Neural Network (RNN) has the ability to model sequential signals and to deal with variable-length utterances. LSTM \cite{hochreiter1997long} and Gated Recurrent Unit (GRU) \cite{cho2014properties} are well-established  ``gating'' models that tackle the problem of gradient vanishing in vanilla RNN. The GRU is able to achieve comparable performance
to LSTM, with a more simplified architecture. It has been applied to utterance-based classification tasks, e.g., question detection \cite{tang2016question} and emotion classification \cite{rana2016gated}.
On the other hand, Convolutional Neural Network (CNN) based approaches have been investigated in the speech assessment area. It was successfully
applied to spoken fluency scoring \cite{chung2017spoken} and differentiate speech of people with Parkinson disease from that of healthy people \cite{vasquez2017convolutional}. In this paper, we focus on investigating the effectiveness of GRU-RNN model and CNN model in PWA speech assessment.
\par Although the CNN-based models are demonstrated to be effective in speech and audio classification tasks, there have been few attempts to provide interpretations of what types of speech or audio patterns are perceived by a CNN model. In the image classification domain, similar issues have been extensively explored. The Class Activation Mapping (CAM) \cite{zhou2016learning} was proposed to highlight the class-discriminative image regions in CNNs with global average pooling. A generalized version of CAM named the Gradient-weighted Class Activation Mapping (Grad-CAM) was proposed in \cite{selvaraju2017grad}, which can be applied to a wider variety of CNN structures. Recently, the Grad-CAM technique was used to analyze the classification results in the audio scene classification task \cite{wu2019enhancing}. Inspired by this, in the present study, we perform the visualization of CAM based on the time-frequency representations of acoustic features, which can be regarded as $2$-dimensional images. It allows the comparison between machine perception and human interpretation, as well as the analysis of what acoustic clues are actually important for deciding the severity degree of PWA.
% ------------------------------------------------------------------------
% ------------------------------------------------------------------------
\section{Corpus: Cantonese AphasiaBank}
\label{sec:dataset}
Cantonese is an influential Chinese dialect spoken by tens of million of people in Hong Kong, Macau, Southern China as well as overseas Chinese communities. Like Mandarin, Cantonese is a monosyllabic and tone language. Each Chinese character is spoken as a monosyllable carrying a specific tone.
\par Cantonese AphasiaBank used in this study is a large-scale multi-modal corpus
jointly developed by the University of Central Florida and the University of Hong Kong \cite{kong2018cantonese}. It aims to support both fundamental and clinical research on Cantonese-speaking aphasia population. The corpus contains recordings of spontaneous speech from $149$ unimpaired and $104$ aphasic subjects who are all native Cantonese speakers. The speech recordings were elicited following the AphasiaBank protocol, with adaptation to local Chinese culture \cite{macwhinney2011aphasiabank,kong2015coding}. Each subject was required to complete $9$ narrative tasks, including $4$ picture descriptions, $1$ procedure description, $2$ story telling and $2$ personal monologues, with details given as in Table \ref{table:tasks}. Except personal monologues, the speech produced in each task is expected to be about a specific topic. All impaired subjects in the corpus went through a standardized assessment using the Cantonese Aphasia Battery \cite{Yiu1992}. It involves a series of sub-tests measuring fluency, information content, comprehension, repetition and naming abilities of the subject. The sum of sub-test scores is named the Aphasia Quotient (AQ). The value of AQ ranging from $0$ to $100$ is an indication of overall severity of impairment. Lower AQ value means higher degree of severity.
\par
\begin{figure}[!htb]
\centering
%  \includegraphics{ISCSLP3.eps}
  	\includegraphics[width=0.7\linewidth]{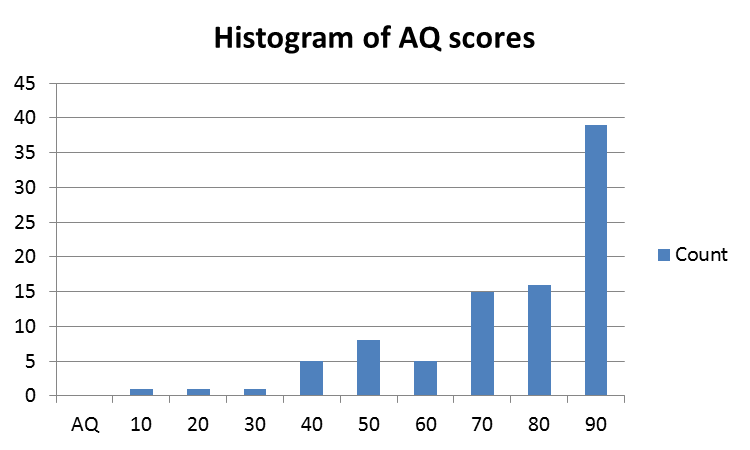}
\caption{Histogram of AQ scores of $91$ impaired speakers.}\label{fig:his}      % Give a unique label
\end{figure}

 In this study, about $15.6$ hours speech data of $9$ tasks from $91$ impaired subjects are selected for the following experiments, including $58$ Anomic subjects, $6$ Transcortical sensory subjects, $12$ Transcortical motor subjects, $10$ Broca’s subjects, $1$ Isolation subject, $2$ Wernicke’s subjects and $2$ Global aphasia subjects. Their AQ values are in the range of $11.0$ to $99.0$.  Figure \ref{fig:his} shows the histogram of the AQ scores of $91$ aphasic subjects. It can be seen that the number of impaired speakers who have relatively high AQ scores is larger than that of those with low AQ scores.
\begin{table*}[]
\centering
\caption{Recording tasks in the Cantonese AphasiaBank.}
\label{table:tasks}
\begin{tabular}{lll}
\hline\noalign{\smallskip}
Task & Recording & Description \\ \noalign{\smallskip}\hline\noalign{\smallskip}
Single picture description & CatRe & Black and white drawing of a cat on a tree being rescued. \\ \noalign{\smallskip}\cline{2-3}\noalign{\smallskip}
 & Flood & A color photo showing a fireman rescuing a girl. \\ \noalign{\smallskip}\hline\noalign{\smallskip}
Sequential picture description & BroWn & Black and white drawing of a boy accidentally break a window. \\ \noalign{\smallskip}\cline{2-3}\noalign{\smallskip}
 & RefUm & Black and white drawing of a boy refusing an umbrella from his mother. \\ \noalign{\smallskip}\hline\noalign{\smallskip}
Procedure description & EggHm & Procedures of preparing a sandwich with egg, ham and bread. \\ \noalign{\smallskip}\hline\noalign{\smallskip}
Story telling & CryWf & Telling a story from a picture book ``The boy who cried wolf''. \\ \noalign{\smallskip}\cline{2-3}\noalign{\smallskip}
 & TorHa & Telling a story from a picture book ``The tortoise and the hare''. \\ \noalign{\smallskip}\hline\noalign{\smallskip}
Personal monologue & ImpEv & Description of an important event in life. \\ \noalign{\smallskip}\cline{2-3}\noalign{\smallskip}
 & Stroke & Description of the experience of suffering a stroke. \\ \noalign{\smallskip}\hline\noalign{\smallskip}
\end{tabular}
\end{table*}
%
% ------------------------------------------------------------------------
% ------------------------------------------------------------------------
\section{General Framework}
\label{sec:general_framework}
\begin{figure}
\centering
%  \includegraphics{ISCSLP3.eps}
  	\includegraphics[width=0.45\linewidth]{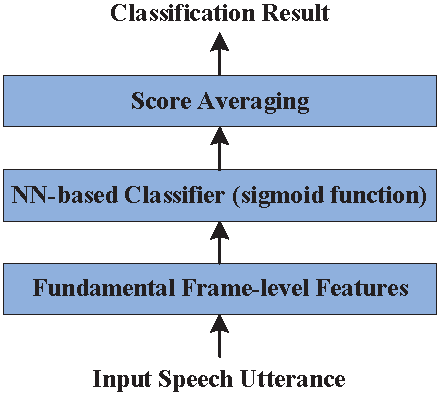}
\caption{General framework of classification system.}\label{fig:framework}
\end{figure}

We propose to utilize end-to-end approach to differentiating PWA with High-AQ (AQ $\geq 90$) from those with Low-AQ (AQ $< 90$). The cutoff value of $90$ is set to reach balanced number of subjects in two groups. Figure \ref{fig:framework} illustrates the general framework of proposed classification system. Fundamental frame-level acoustic features (e.g. Mel-frequency cepstral coefficients, filterbank features) are extracted from the utterance of PWA and subsequently fed to a Neural Network-based (NN-based) classifier. Under the ``end-to-end'' framework, neither  pathology-specific feature extraction nor feature selection procedures are required. They are expected to be automatically carried out by the NN-based classifiers. The classification label of the input utterance is inherited from the impaired speaker. For the output of classifier, an assessment score for each utterance can be obtained after the sigmoid function of the neural network. The speaker-level overall assessment score is computed by taking the average of all utterance-level scores from the test speaker. The higher overall assessment score indicates a higher possibility of the speaker being in the High-AQ group.
%
% ------------------------------------------------------------------------
% ------------------------------------------------------------------------
\section{Methods}
\label{sec:Methods}
\subsection{Models}
In this study, we experiment with two model structures, namely the sequence-to-one GRU-RNN and CNN for the classification purpose.
% ------------------------------------------------------------------------
\subsubsection{Sequence-to-one GRU-RNN}
\begin{figure}
\centering
%  \includegraphics{ISCSLP3.eps}
  	\includegraphics[width=0.65\linewidth]{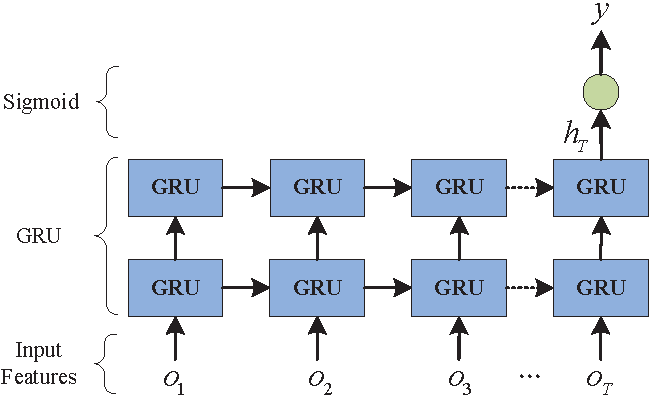}
\caption{Architecture of sequence-to-one GRU-RNN model.}
\label{fig:GRU-RNN}       % Give a unique label
\end{figure}

GRU-RNN is specialized in sequential modeling, which is expected to characterize the impairment of the entire utterance spoken from PWA. The architecture of proposed sequence-to-one GRU-RNN model is detailed in Figure \ref{fig:GRU-RNN}. Frame-level acoustic features $\bm{o}_1, \bm{o}_2, \cdots, \bm{o}_T$ of an utterance are sequentially fed to the neural network. A hidden representation is generated at each time step, summarizing the past acoustic information. The hidden representation $\bm{h}$ at the $t^{th}$ time step is given by
\begin{equation}
    \bm{h}_t = \text{GRU}(\bm{o}_1, \bm{o}_2, \cdots, \bm{o}_t; \bm{\theta}_h),
\end{equation}
where $\bm{\theta}_h$ denotes the model parameters and the function GRU($\cdot$) indicates two uni-directional GRU layers with hidden size of $200$ per layer. Our preliminary experiments show no significant improvement using the bi-directional GRU layers, thus a uni-directional structure is adopted in this study. A predicted score $y$ for the input utterance is produced by a sigmoid activation function Sigmoid($\cdot$) at the output layer, which is given by
\begin{equation}
    y = \text{Sigmoid}(\bm{W}_o \bm{h}_T+\bm{b}_o),
\end{equation}
where $\bm{h}_T$ is the hidden representation at the last time step which summarizes the information of entire utterance. $\bm{W}_o$ and $\bm{b}_o$ are the trainable parameters connecting $\bm{h}_T$ with $y$.
% ------------------------------------------------------------------------
\subsubsection{Convolutional Neural Network}
CNN is known as to learn local high-level features from the spatial or temporal data. It performs discrete convolution between a local region of the input image $f$ and a set of filters $w$ with the shapes of $m\times n$. After the input image $f$ passes through the convolutional layer, the output is given by
\begin{equation}
    (f * w)[i,j]=\sum_m\sum_n f[i+m,i+n]w[m,n].
\end{equation}
For the input utterance, the short-time acoustic features are extracted and concatenated sequentially as a $2$-dimensional time-frequency representation. The CNN model structure described in Table \ref{table:CNN} is inspired by the AlexNet \cite{krizhevsky2012imagenet} and VGG \cite{simonyan2014very} models. It is constructed with a Global Average Pooling (GAP) layer instead of a fully-connected layer after the last convolutional layer. GAP has been proved to be a good regularizer for CNNs in image classification \cite{lin2013network} and audio scene classification \cite{wu2019enhancing} tasks. Batch normalization (BN) layer and ReLU activation function are added after each convolutional layer.
\begin{table}
\centering
\caption{Architecture of the CNN model.}
\label{table:CNN}      % Give a unique label
\begin{tabular}{cc}
\hline\noalign{\smallskip}
 & Input 1$\times$300$\times$128 \\ \noalign{\smallskip} \hline\noalign{\smallskip}
1 & 3$\times$3 Convolution (pad-1, stride-1)-64-BN-ReLU \\ \noalign{\smallskip}
2 & 3$\times$3 Max Pooling (stride-2) \\ \noalign{\smallskip} \hline\noalign{\smallskip}
3 & 3$\times$3 Convolution (pad-1, stride-1)-192-BN-ReLU \\ \noalign{\smallskip}
4 & 3$\times$3 Max Pooling (stride-2) \\ \noalign{\smallskip} \hline\noalign{\smallskip}
5 & 3$\times$3 Convolution (pad-1, stride-1)-384-BN-ReLU \\ \noalign{\smallskip}
6 & 3$\times$3 Convolution (pad-1, stride-1)-256-BN-ReLU \\ \noalign{\smallskip}
7 & 3$\times$3 Convolution (pad-1, stride-1)-256-BN-ReLU \\ \noalign{\smallskip}
8 & 3$\times$3 Max Pooling (stride-2) \\ \noalign{\smallskip} \hline\noalign{\smallskip}
9 & Global Average Pooling \\ \noalign{\smallskip}
10 & Sigmoid \\
\noalign{\smallskip}\hline
\end{tabular}
\end{table}
% ------------------------------------------------------------------------
\subsection{Class Activation Mapping}\label{subsec:CAM}
Given a CNN model, the technique of Class Activation Mapping (CAM) \cite{zhou2016learning} is capable of visualizing the class-discriminative regions in the input images. Therefore, the CAM can improve understanding and interpretability of CNN models. Given a trained CNN model with GAP, the mean values of $K$ feature maps at the penultimate layer are linearly transformed to generate a score $y^c$ for each class $c$,
\begin{equation}\label{eq:1}
    y^c = \sum_k w_k^c ~\frac{1}{Z}\sum_{x}\sum_{y}f_k(x, y),
\end{equation}
where $f_k(x, y)$ denotes the point $(x, y)$ in the $k^{th}$ feature map before GAP. The weight $w_k^c$ connecting $k^{th}$ feature map to output class $c$ indicates the importance of $k^{th}$ feature map for the target class $c$. $Z$ is a normalization term, representing the number of pixels in the feature map. The spatial elements of class activation map $L_{\text CAM}^c$ for class $c$ are obtained by exchanging the order of summation in equation \eqref{eq:1},
\begin{equation}\label{eq:2}
   L_{\text{CAM}}^c(x, y) = \sum_k w_k^cf_k(x, y).
\end{equation}
\par The above CAM is only applicable to CNNs with GAP. As a strict generalization of CAM, the Gradient-weighted Class Activation Mapping (Grad-CAM) can be applied to a wider variety of CNN models, such as those with fully-connected layers (e.g. VGG, AlexNet). It is implemented by replacing the weight of each feature map $w_k^c$ with the average gradient ($\alpha_k^c$) of the score for class $c$ with respect to each feature map, which is computed by
\begin{equation}\label{eq:3}
   \alpha_k^c = \frac{1}{Z} \sum_x\sum_y\frac{\partial y^c}{\partial f_k(x, y)}.
\end{equation}
Then we have
\begin{equation}\label{eq:4}
   L^c_{\text{Grad-CAM}} (x, y) = \text{ReLU}(\sum_k \alpha_k^cf_k(x, y)),
\end{equation}
where ReLU is applied to identify the features that have a positive influence on the target class. Note that $f_k$ in this case is no longer limited to the last convolutional layer in CNN but can be from any convolutional layer.
\par We propose to use Grad-CAM technique to analyze a trained CNN model for the assessment task. It is expected to help the understanding of the feature patterns learned by CNN in the decision of severity degree. Note that the class number $c$ is equal to $1$ and the sigmoid function is used in this study. Equation \eqref{eq:4} is used to highlight the positive regions to the High-AQ class. By negating the weighted combination before the ReLU function, the negative regions to the High-AQ class (positive to Low-AQ class) could be visualized.
% ------------------------------------------------------------------------
% ------------------------------------------------------------------------
\section{Experimental Setup}
% ------------------------------------------------------------------------
\subsection{Data}
To perform severity assessment for all $91$ PWA, the binary classification experiment is carried out with the arrangement of $5$-fold cross validation. In each fold, $80\%$ of the subjects are used for training and the rest $20\%$ subjects are used for test. $10\%$ subjects are randomly selected from training subjects as the validation data. There are $39$ PWA in High-AQ group and $52$ PWA in Low-AQ group respectively.
\par We consider to use two types of segmentation methods to process speech recordings. The first method is manual segmentation according to the sentence boundaries marked in the Cantonese AphasiaBank. In this way, the speech recordings of $91$ PWA are cut into $11,550$ utterances. A total number of $6, 172$ utterances spoken from High-AQ speakers are assigned with label $1$, while $5, 378$ utterances from Low-AQ group are assigned with label $0$. The lengths of utterances range from $0.07$ second to $70.03$ seconds. Table \ref{table:number_GRU} details the $5$-fold data sets with the manual segmentation. The second segmentation method is to cut the speech recordings of $91$ PWA into non-overlapping segments of $3$ seconds, leading to $18, 149$ equal-length short utterances for training models. There are $7, 322$ utterances and $10, 827$ utterances being labeled as $1$ and $0$ respectively. The number of equal-length utterances in $5$ folds is listed in Table \ref{table:num_CNN}.
\begin{table}
\centering
\caption{The number of variable-length utterances in each fold by manual segmentation.}
\label{table:number_GRU}      % Give a unique label
\begin{tabular}{lccccc}
\hline\noalign{\smallskip}
 & Fold 1 & Fold 2 & Fold 3 & Fold 4 & Fold 5 \\ \noalign{\smallskip}\hline\noalign{\smallskip}
Train & $8000$ & $8278$ & $8207$ & $8535$ & $8368$ \\ \noalign{\smallskip}
Test & $2623$ & $2001$ & $2389$ & $2261$ & $2276$ \\ \noalign{\smallskip}
Valid & $927$ & $1271$ & $954$ & $754$ & $906$ \\ \noalign{\smallskip}
\hline
\end{tabular}
\end{table}
\begin{table}
\centering
\caption{The number of equal-length utterances of $3$ seconds in each fold.}
\label{table:num_CNN}
\begin{tabular}{lccccc}
\hline\noalign{\smallskip}
 & Fold 1 & Fold 2 & Fold 3 & Fold 4 & Fold 5 \\ \noalign{\smallskip}\hline\noalign{\smallskip}
Train & $12839$ & $13270$ & $13155$ & $13984$ & $12669$ \\ \noalign{\smallskip}
Test & $4100$ & $3224$ & $3725$ & $2940$ & $4160$ \\ \noalign{\smallskip}
Valid & $1210$ & $1655$ & $1269$ & $1225$ & $1320$ \\ \noalign{\smallskip}
\hline
\end{tabular}
\end{table}
% ------------------------------------------------------------------------
\subsection{Features}
Mel-Frequency Cepstral Coefficient (MFCC) and  Log-Mel filterbank features are explored to train the proposed models. The $39$-dimensional MFCC$+\Delta+\Delta\Delta$ features are extracted from each input utterance, with $25$ milliseconds window length and $10$ milliseconds window shift. The number of points in Discrete Fourier Transform (DFT) is set to $2048$, and the number of filters in the Mel  filterbank is set to $128$. Dimension-wise normalization of the MFCC features is performed within each fold using the mean and
variance computed from the training set. For the extraction of Log-Mel filterbank features, the settings of DFT and Mel filterbank are the same as those used for extracting the MFCC features. The Log-Mel filterbank features are derived from short-time frames (same window size and window shift as mentioned before) from input utterances, with the dimension of $128$. Subsequently, the frame-level features are stacked together as a time-frequency matrix for the input utterance. The feature normalization is also applied to Log-Mel filterbank features in each fold.
\par In addition, topic information is taken into consideration in the feature design. With the topic information, the content-related acoustic patterns are expected to be learned by the NN-based models.  For instance, if an utterance is from the topic $1$, the first element of a topic vector (topicvec) is set to $1$ and other $8$ elements are set to zeros. For training the sequence-to-one GRU-RNN model, the  $9$-dimensional one-hot topic vector is concatenated with a $128$-dimensional Log-Mel filterbank feature, resulting in a $137$-dimensional vector for each frame as a kind of topic adaptation. For the CNN model, the one-hot topic vectors are utilized in a different way. The topic vector is concatenated with the embedding of time-frequency representation of the utterance. Specifically, the $256$-dimensional high-level features derived from the Log-Mel filterbank features can be obtained after the GAP layer, which is shown in Table \ref{table:CNN}. They are concatenated with $9$-dimensional one-hot topic vectors and then fed to a fully-connected neural network containing two hidden layers (the vector size of each hidden layer is $256$). It is followed by a sigmoid layer as the final output layer.
\par In the following experiments, we consider to use MFCC, Log-Mel filterbank and topic-adapted features derived from two types of segmented speech data to train the sequence-to-one GRU-RNN model and CNN model.
% ------------------------------------------------------------------------
\subsection{Hyperparameters for Model Training}
The training parameters are set empirically. The mini-batch size is $64$ for training the sequence-to-one GRU-RNN model and the CNN model. The initial learning rate is set to $10^{-4}$ for the sequence-to-one GRU-RNN model and it is set to $10^{-3}$ for the CNN model. Model training aims at minimizing the binary cross-entropy loss using the Adam
optimizer \cite{kingma2014adam}. Weight decay coefficient is set to $5\times10^{-4}$ for both models to tackle the overfitting problem. All models are implemented using the deep learning toolkit Pytorch \cite{paszke2017automatic}.
% ------------------------------------------------------------------------
% ------------------------------------------------------------------------
\section{Experiments and Results}
% ------------------------------------------------------------------------
\subsection{Utterance-level Classification Accuracy}
For the binary classification on test utterances from High-AQ and Low-AQ subjects, the Area Under receiver operating characteristic Curve (AUC) \cite{vuk2006roc} is adopted as the performance metric. In the case of binary classification, AUC can be viewed as the probability that the classifier ranks a randomly chosen positive sample higher than a negative one \cite{fawcett2006introduction}. An AUC value $0.5$ means a random guess and $1.0$ represents a perfect classification.
% \usepackage{multirow}
\begin{table}
\centering
\caption{AUC result of test data in each fold based on sequence-to-one GRU-RNN (GRU-RNN) and CNN models. MFCC, Log-Mel and topic-adapted features derived from variable-length (-var) and equal-length (-eq) utterances are used for training the models.}
\label{table:AUC}
%\resizebox{\textwidth}{!}{%
\begin{tabular}{llccccc}
\hline\noalign{\smallskip}
{Model} & {Feature} & \multicolumn{5}{c}{AUC} \\ \noalign{\smallskip} \cline{3-7}\noalign{\smallskip} \noalign{\smallskip}
 &  & Fold 1 & Fold 2 & Fold 3 & Fold 4 & Fold 5 \\  \noalign{\smallskip}\hline\noalign{\smallskip}
GRU-RNN & MFCC-var & $0.58$ & $0.62$ & $0.58$ & $0.61$ & $0.80$ \\ \noalign{\smallskip}
& Log-Mel-var & $0.62$ & $0.65$ & $0.62$ & $0.61$ & $0.80$ \\ \noalign{\smallskip}
  & Log-Mel-eq & $0.62$ & $0.69$ & $0.64$ & $0.61$ & $0.81$ \\ \noalign{\smallskip}
   & Log-Mel+topicvec-eq & $0.64$ & $0.75$ & $0.65$ & $0.63$ & $0.83$ \\ \noalign{\smallskip}
CNN & Log-Mel-eq & \bm{$0.74$} & 0.81 & \bm{$0.78$} & \bm{$0.70$} & \bm{$0.90$} \\ \noalign{\smallskip}
 & Log-Mel+topicvec-eq & 0.70 & \bm{$0.82$} & 0.76 & \bm{$0.70$} & 0.89 \\ \noalign{\smallskip}
\hline
\end{tabular}
%\end{tabular}%
%}
\end{table}
\par Table \ref{table:AUC} compares the AUC results of $5$-fold cross-validation experiments with sequence-to-one GRU-RNN model (GRU-RNN model) and CNN model. The ``-var'' and ``-eq'' denote the acoustic features  extracted from variable-length and equal-length utterances respectively.
From the AUC results based on the GRU-RNN model, it can be seen that higher AUCs are obtained using the Log-Mel filterbank (Log-Mel) features than using the MFCC features. This suggests that the Log-Mel features are more suitable for the intended classification task. The result comparison between Log-Mel-var and Log-Mel-eq shows that the equal-length utterances perform better than the variable-length ones for training the GRU-RNN model. It may attribute to a larger number of utterances used for training.
With the Log-Mel-eq features, a significant AUC improvement can be observed using the CNN model compared with using the GRU-RNN model.
This probably because the CNN model performs better on capturing localized time-frequency information of impairment-related acoustic patterns.
Nevertheless, the CNN model does not benefit much from the topic vectors for the severity degree classification, whilst for the sequence-to-one GRU-RNN model, using the concatenation of Log-Mel and topic vector features achieves better performance than using the Log-Mel features only. This shows that the topic information may provide additional benefits to the classification performance to some extent.
\par The sequence-to-one GRU-RNN model trained with Log-Mel+topicvec-eq features attains the best performance among all sequence-to-one GRU-RNN models, thus it is used for the analysis in section \ref{subsec:speaker acc} and section \ref{sec:gruvscnn}. For the CNN model, it is noted that the topic vectors show little improvement on the classification performance. Therefore, the CNN model trained with Log-Mel-eq without topic vectors is used for the following analysis of experimental results and Grad-CAM visualization.
\par Although the utterance-level AUC is a direct metric to measure the performance of classifiers, it is unable to reflect the performance of severity degree assessment for PWA. A score fusion procedure is required to give a speaker-level classification decision.
% ------------------------------------------------------------------------
\subsection{Speaker-level Classification Accuracy}\label{subsec:speaker acc}
We consider to compare the speaker-level classification accuracy given by the sequence-to-one GRU-RNN model with the best performance (trained with Log-Mel+topicvec-eq features), the CNN model (trained with Log-Mel-eq features) and a baseline system. As shown in Figure \ref{fig:framework}, for the NN-based models, after obtaining utterance-level scores from a test speaker, a score fusion is performed to combine them as an overall score for the speaker-level classification. The overall score for each test speaker is obtained by taking the average of speaker's utterance-level scores. The threshold for binary classification is set to $0.5$. If the overall score is higher than $0.5$, the test speaker is classified as High-AQ, otherwise the speaker is classified as Low-AQ.
\par The baseline assessment system in this study follows a conventional two-step assessment approach proposed in our previous study \cite{Qin2018icassp}. A $5$-dimensional feature vector of supra-segmental duration features is evaluated on the same task of binary classification using a random forest classifier. The leave-one-out cross validation  strategy is adopted. The feature vector includes: (1) duration ratio between non-speech part and speech part; (2) average duration of silence segments (longer than $0.5$ second); (3) average duration of speech segments (speech region between two silence segments); (4) ratio of silence segment count to syllable count; (5) syllable count per second. All of the features are generated from the time alignment of a dedicated ASR system. The time-delay layers stacked with bidirectional long short term memory layers (TDNN-BLSTM) are used as  acoustic model of the ASR system and it is trained using multi-task learning strategy \cite{Qin2018IS}. These ASR-generated features were shown to be effective to classify High-AQ speakers from Low-AQ ones in the aspect of acoustic impairment of PWA speech \cite{Qin2018icassp}.
% \usepackage{graphicx}
\begin{table}
\centering
\caption{Classification performance of baseline system, sequence-to-one GRU-RNN model and CNN model on $91$ test speakers. It is measured in terms of the accuracy, average F1 score, recall and specificity.}
\label{table:acc}
%\resizebox{\textwidth}{!}{%
\begin{tabular}{lccc}
\hline\noalign{\smallskip}
Model & Accuracy & F1 & Recall/Specificity \\ \noalign{\smallskip}\hline\noalign{\smallskip}
Baseline & $0.813$ & $0.810$ & $0.795$/$0.827$ \\\noalign{\smallskip}
GRU-RNN & $0.791$ & $0.789$ & $0.795$/$0.789$ \\\noalign{\smallskip}
CNN & \bm{$0.824$} & \bm{$0.822$} & \bm{$0.821$}/\bm{$0.827$} \\ \noalign{\smallskip}\hline
%\end{tabular}%
%}
\end{tabular}
\end{table}
\par Table \ref{table:acc} lists the speaker-level binary classification results on $91$ test speakers. The classification performance is measured in terms of the accuracy, average F1 score, recall and specificity. It is seen that the CNN model performs the best among the three models, while the performance of sequence-to-one GRU-RNN model is worse than that of the baseline system. This demonstrates that the proposed CNN model is preferred than the sequence-to-one GRU-RNN model for this specific assessment task. The CNN model outperforms the baseline system in terms of the recall, with the value of $0.821$ $(32/39)$ versus $0.795$ $(31/39)$. Table \ref{table:confusion} shows the confusion matrix of classification result using the CNN model. The result suggests that the end-to-end approach has great potential to replace the traditional two-step method to perform speech assessment for PWA.
% \usepackage{graphicx}
\begin{table}
\centering
\caption{Confusion matrix of classification result using CNN model.}
\label{table:confusion}
%\resizebox{\textwidth}{!}{%
\begin{tabular}{lccc}
\hline\noalign{\smallskip}
 & \multicolumn{1}{l}{True: High-AQ} & \multicolumn{1}{l}{True: Low-AQ} & \multicolumn{1}{l}{Total} \\ \noalign{\smallskip}\hline\noalign{\smallskip}
Predicted: High-AQ & $32$ & $9$ & $41$ \\\noalign{\smallskip}
Predicted: Low-AQ & $7$ & $43$ & $50$ \\ \noalign{\smallskip}
Total & $39$ & $52$ & $91$ \\ \noalign{\smallskip}\hline
%\end{tabular}%
%}
\end{tabular}
\end{table}
% ------------------------------------------------------------------------
% ------------------------------------------------------------------------
\section{Discussion}
% ------------------------------------------------------------------------
\subsection{Sequence-to-one GRU-RNN vs. CNN}\label{sec:gruvscnn}
\begin{figure}[!htb]
\begin{minipage}[b]{.48\linewidth}
  \centering
  \centerline{\includegraphics[width=6.8 cm]{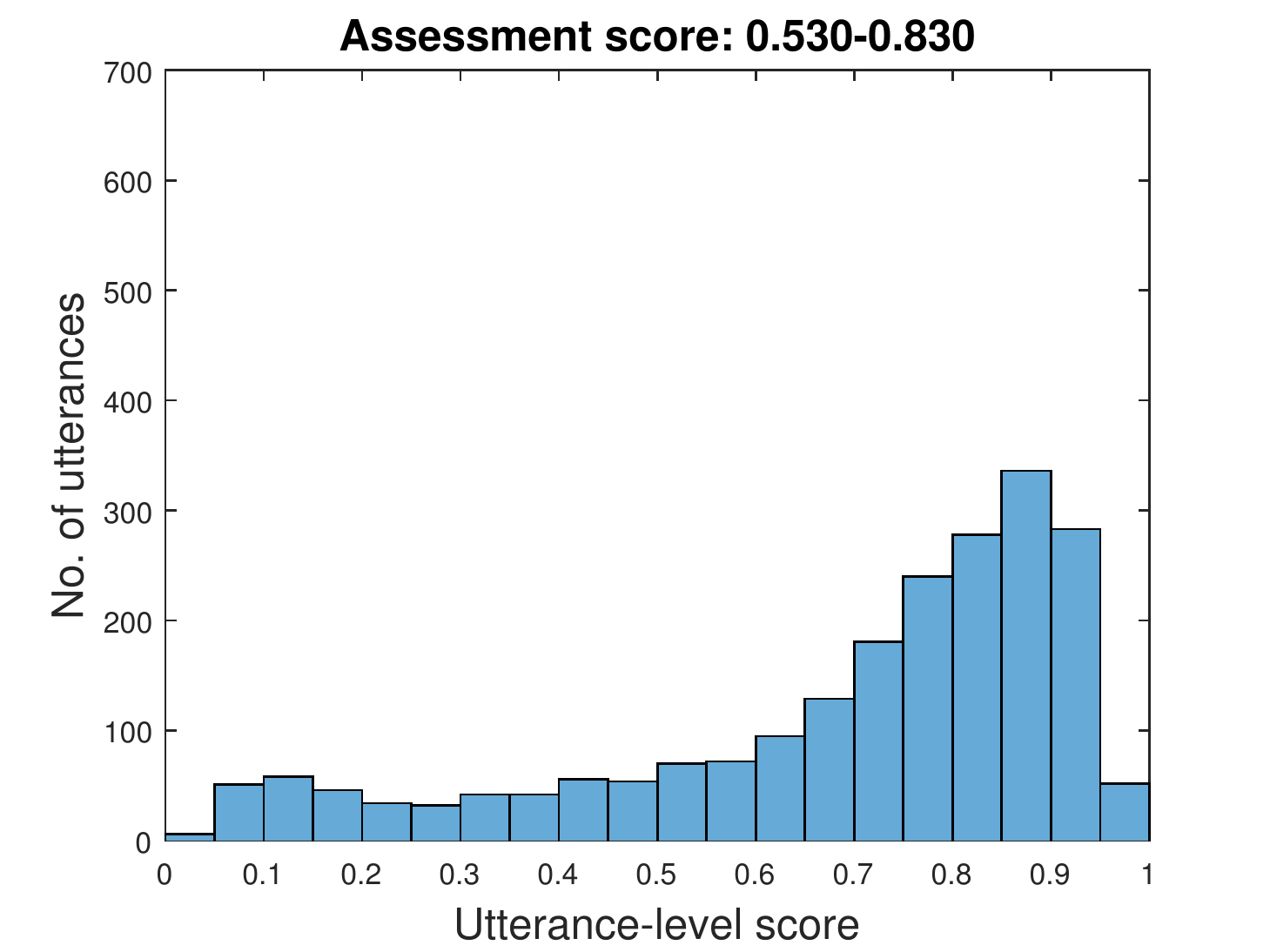}}
 %\vspace{0.02cm}
  \centerline{(a) Sequence-to-one GRU-RNN}%\smallskip
\end{minipage}
\hfill
\begin{minipage}[b]{0.48\linewidth}
  \centering
  \centerline{\includegraphics[width=6.8 cm]{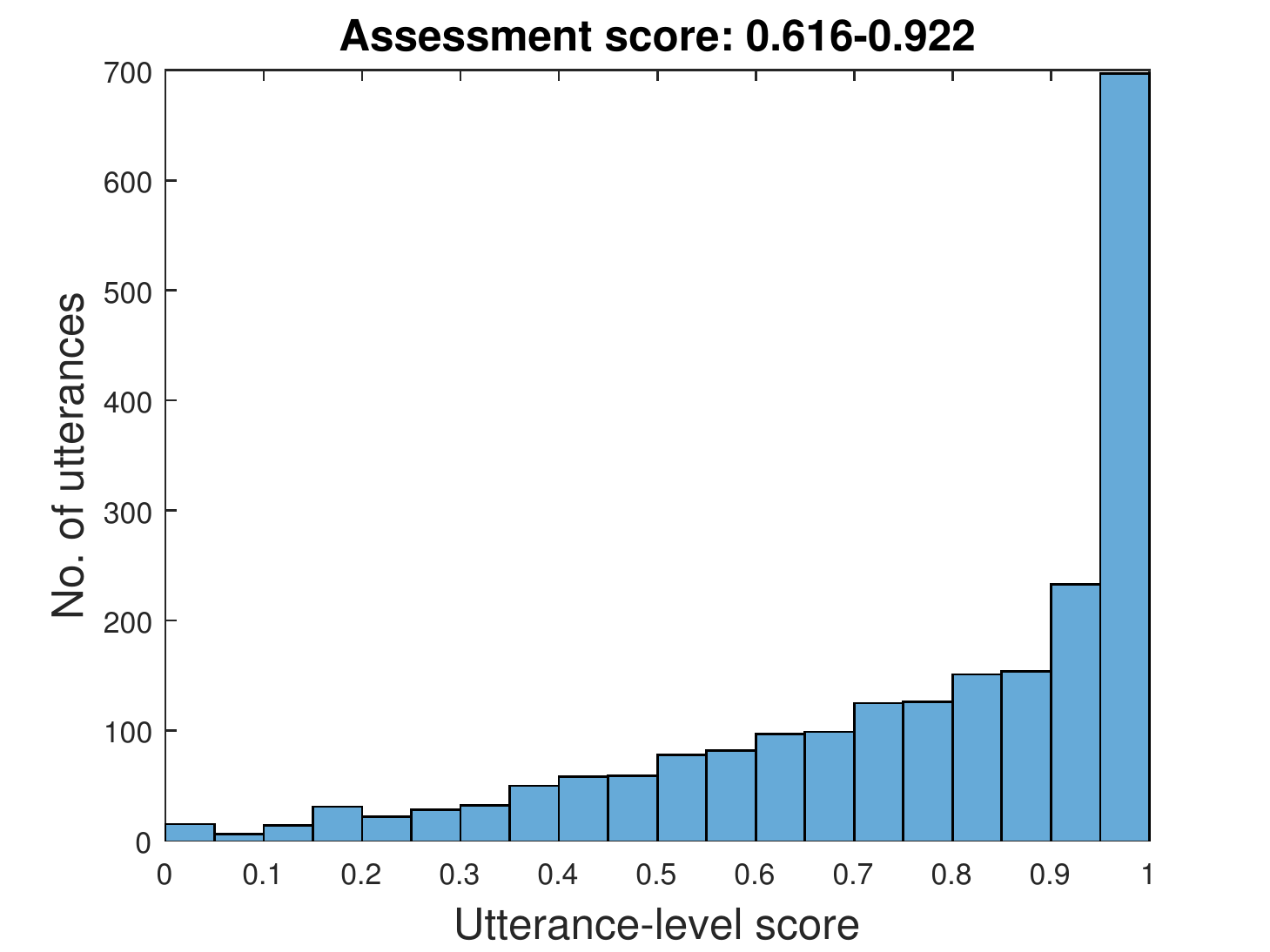}}
  %\vspace{0.02cm}
  \centerline{(b) CNN}%\smallskip
\end{minipage}
\caption{Comparison between histogram of utterance-level scores from sequence-to-one GRU-RNN model and that from CNN model for ten speakers with High-AQ.}
\label{fig:his-1}
\end{figure}

\begin{figure}[!htb]
\begin{minipage}[b]{.48\linewidth}
  \centering
  \centerline{\includegraphics[width=6.8 cm]{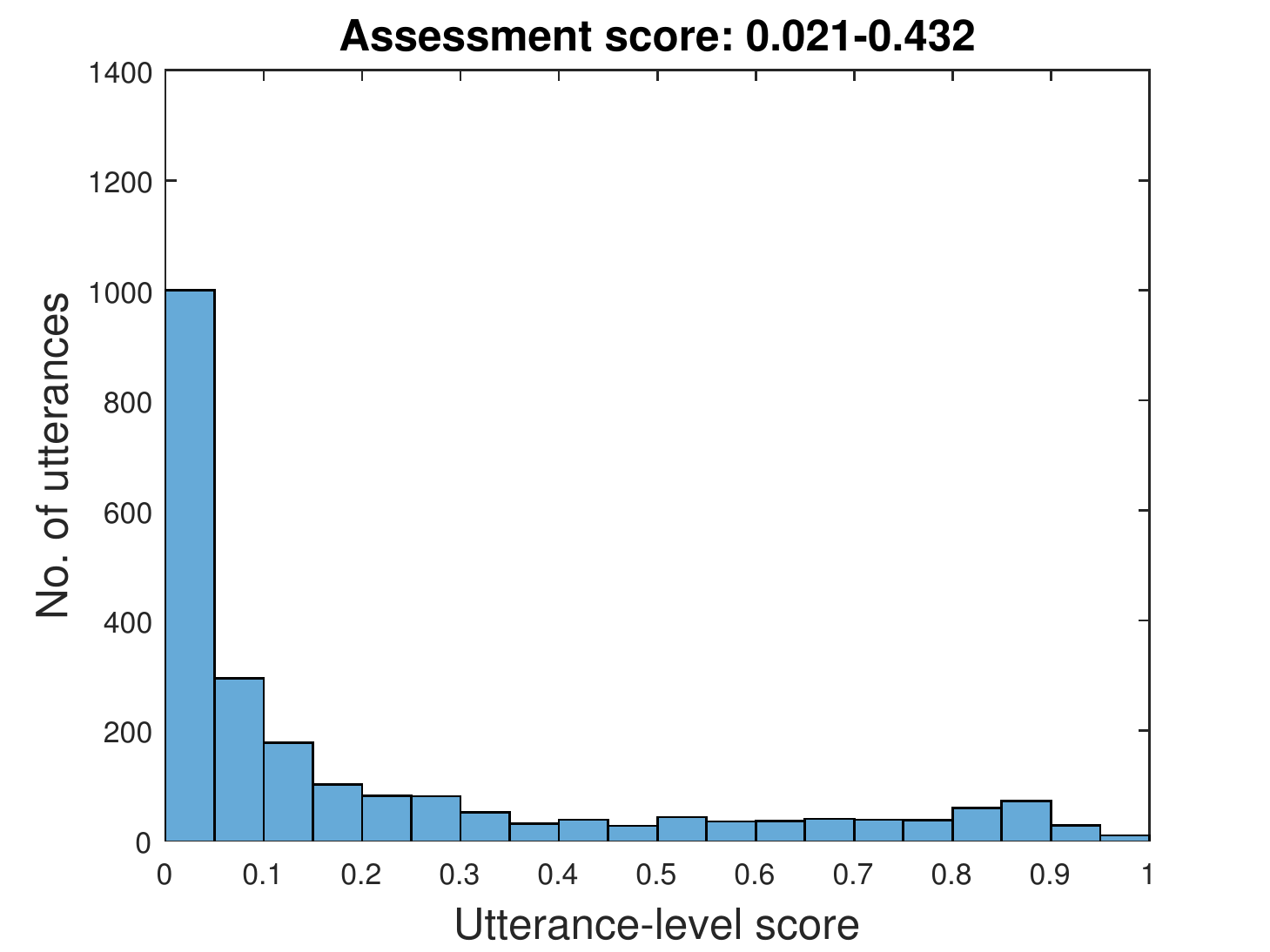}}
 %\vspace{0.02cm}
  \centerline{(a) Sequence-to-one GRU-RNN}%\smallskip
\end{minipage}
\hfill
\begin{minipage}[b]{0.48\linewidth}
  \centering
  \centerline{\includegraphics[width=6.8 cm]{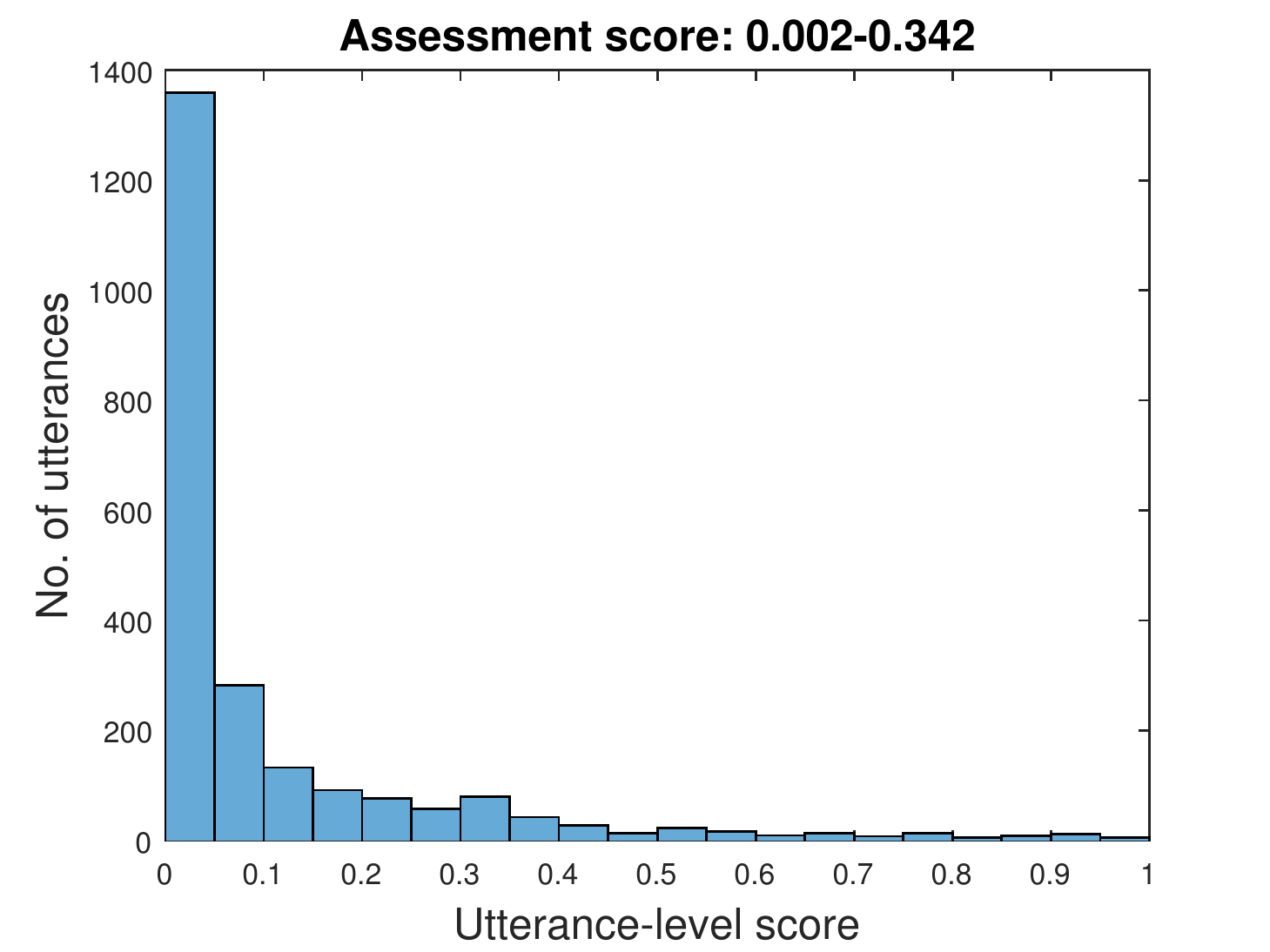}}
  %\vspace{0.02cm}
  \centerline{(b) CNN}%\smallskip
\end{minipage}
\caption{Comparison between histogram of utterance-level scores from sequence-to-one GRU-RNN model and that from CNN model for ten speakers with Low-AQ.}
\label{fig:his-0}
\end{figure}

To investigate the performance of the sequence-to-one GRU-RNN and CNN models, we analyze twenty typical impaired subjects for whom the classification results are correct. Ten of the selected speakers are with High-AQ while the others are with Low-AQ. Figure \ref{fig:his-1} illustrates the histograms of utterance-level scores given by the sequence-to-one GRU-RNN and CNN for High-AQ speakers respectively. Although two models make correct decisions on these speakers, their overall assessment scores generated from the CNN (range from $0.616$ to $0.922$) is much nearer to $1.0$ than those from the sequence-to-one GRU-RNN (range from $0.530$ to $0.830$). Specifically, most of the utterance-level scores are higher than $0.9$ with the CNN model, while those are mainly distributed in the range of $0.7$ to $0.9$ based on the sequence-to-one GRU-RNN model. Similar to the previous case, the CNN performs better than the sequence-to-one GRU-RNN in generating overall scores for ten speakers with Low-AQ ($0.002-0.342$ versus $0.021-0.432$). The distribution of utterance-level scores produced by the CNN model is even denser in the low-score region as shown in Figure \ref{fig:his-0}. Generally speaking, utterance-level scores from the CNN model tend to be more polarized than those from the sequence-to-one GRU-RNN model, meaning the CNN model has a higher classification confidence. This is also consistent with the AUC results as shown in Table \ref{table:AUC}.
% ------------------------------------------------------------------------
\subsection{Visualization with Class Activation Maps}
Given a trained CNN model, the Grad-CAM technique is applied to explore what types of features in impaired speech are learned in the CNN classifier to perform assessment. Recall that $3$-second Log-Mel features are as the input to the CNN model, and they can be viewed as time-frequency gray-scale images. After simply up-sampling the class activation maps to the same size of the input image, the transformed class activation maps are further combined with the input image for localizing the class-discriminative regions. The proposed CAM visualization consists of three image components. The base one is the gray-scale Log-Mel image. The second component is a semi-transparent image in red derived from the class activation map. It contains the time-frequency regions that positively influence the classification score of the High-AQ class. For the last component, it is a semi-transparent image of blue color, in which the blue areas indicate regions having a negative influence on the class of High-AQ. The detailed computations of class activation maps for creating the second and third components are described in section \ref{subsec:CAM}.
\par
\begin{figure}[!htb]
\begin{minipage}[b]{1.0\linewidth}\label{fig:4a}
  \centering
  \centerline{\includegraphics[width=\linewidth]{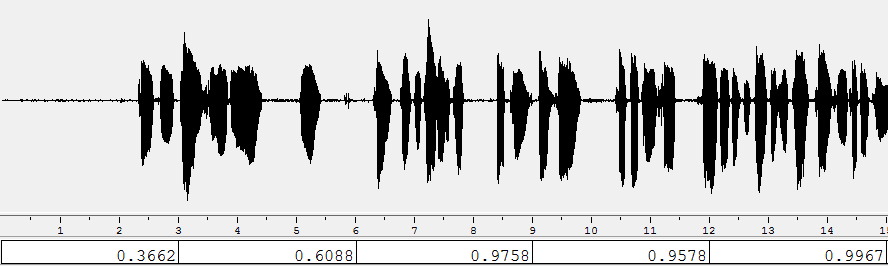}}
%  \vspace{2.0cm}
  \centerline{(a) Utterance-level scores (bottom) of a speech segment with label 1 from CNN model.}\smallskip
\end{minipage}
\begin{minipage}[b]{1.0\linewidth}\label{fig:4b}
  \centering
  \centerline{\includegraphics[ height=2.5cm,width=\linewidth]{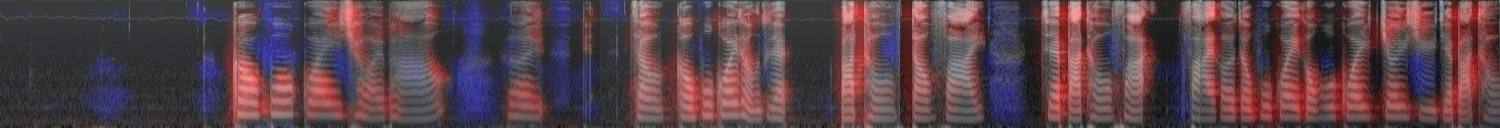}}
%  \vspace{1.5cm}
  \centerline{(b) CAM visualization of a speech segment with label 1 derived from CNN model.}\smallskip
\end{minipage}
\caption{Utterance-level scores and CAM visualization of the same speech segment from a High-AQ speaker. They are derived from the CNN model with Log-Mel-eq input. The positive and negative activation for target class are highlighted in red and blue colours respectively.}
\label{fig:CAM1}
\end{figure}

\begin{figure}[!htb]
\begin{minipage}[b]{1.0\linewidth}
  \centering
  \centerline{\includegraphics[width=\linewidth]{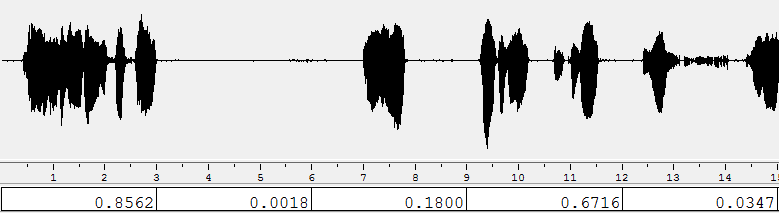}}
%  \vspace{2.0cm}
  \centerline{(a) Utterance-level scores (bottom) of a speech segment with label 0 from CNN model.}\smallskip
\end{minipage}
\begin{minipage}[b]{1.0\linewidth}
  \centering
  \centerline{\includegraphics[height=2.5cm,width=\linewidth]{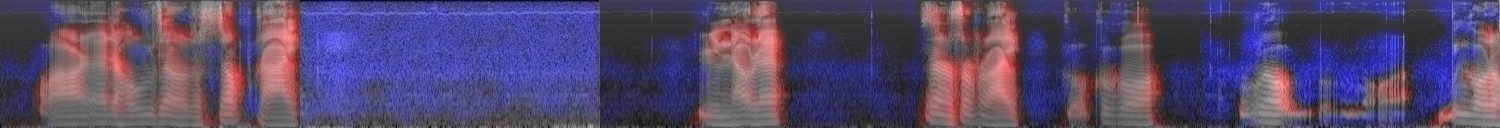}}
%  \vspace{1.5cm}
  \centerline{(b) CAM visualization of a speech segment with label 0 derived from CNN model.}\smallskip
\end{minipage}
\caption{Utterance-level scores and CAM visualization of the same speech segment from a Low-AQ speaker. They are derived from the CNN model with Log-Mel-eq input. The positive and negative activation for target class are highlighted in red and blue colours respectively.}
\label{fig:CAM0}
\end{figure}

 Figure \ref{fig:CAM1} and Figure \ref{fig:CAM0} give utterance-level scores and CAM visualizations of two $15$-second speech segments from a High-AQ speaker (AQ: $94.7$) and a Low-AQ speaker (AQ: $74.2$) respectively. These two speakers are correctly classified by the CNN model. The CAM visualizations are derived from the last convolutional layer ($7^{th}$ layer) in the CNN structure, since the last convolutional layer was expected to keep balance between high-level semantics and detailed spatial information \cite{selvaraju2017grad}. It is clearly observed that the area of positive activation (red) in speech from the High-AQ speaker is larger than that from the Low-AQ speaker. In general, we observe that the CNN model is able to learn the following features to determine the severity degree of PWA:
\begin{enumerate}
    \item  Duration-related features:\\
    As illustrated in Figure \ref{fig:CAM1} (b) and Figure \ref{fig:CAM0} (b), silence parts are usually highlighted as negative activation in blue, while speech parts are as positive activation in red. Meanwhile, 3-second utterances with longer pauses tend to have lower classification scores from CNN model, which can be observed in Figure \ref{fig:CAM1} (a) and Figure \ref{fig:CAM0} (a). This indicates that milder PWA (with higher AQ) tend to exhibit less dysfluency and vice versa. This also confirms that the CNN is able to automatically learn similar duration-related features as we designed in the baseline system, such as ``average duration of silence segments'' and ``average duration of speech segments''.
    \item  Transitions between speech parts and silence parts: \\
    As we can see from Figure \ref{fig:CAM1} (b) and Figure \ref{fig:CAM0} (b), the magnitudes of positive activation (red) are higher at the transition regions between speech parts and silence parts. Thus, speakers who have high speaking rate with a large number of transitions in their speech have a higher probability to be classified as High-AQ. This CNN-learned feature is also matched with one of the conventional features named ``syllable count per second'' used in the baseline system.
    \item  Significant variation of formants: \\
    It can be seen that the positive activation (red) concentrate more on the time-frequency regions with significant variation of formants. Speakers who produce utterances with a variety of formant changes are more likely to be classified into the High-AQ group. This reveals that the fundamental formant information is important to the classification result and it can be implicitly learned by the CNN model. As mentioned in section \ref{subsec:speaker acc}, the formant feature is not considered in the baseline system, which may be the reason of the poorer classification performance than the CNN model. The lack of formant changes could be related to phonetic disorder, voice disorder and limited vocabulary in PWA speech. This suggests that using formant features for the PWA speech assessment is promising and it will be investigated in the future.
\end{enumerate}
Overall, with the CAM visualization, the CNN model is demonstrated to be capable of implicitly learning impaired acoustic patterns for the PWA speech assessment.
%
% ------------------------------------------------------------------------
\subsection{Limitations of Study}
We analyze one test speaker (AQ: $66.8$) who is mis-classified into High-AQ group with the CNN model. It is found that there are a large number of function words and filler words but few topic-specific words in his utterance, even though his utterances are quite fluent. This reveals a major limitation of proposed model: it fails to sufficiently learn the semantic content of utterances but mainly focuses on acoustic impairment of PWA. In the future, we propose to establish another neural network that aims at characterizing language impairment of PWA. A combined neural network considering both acoustic and language impairments will be investigated afterwards.
% ------------------------------------------------------------------------
% ------------------------------------------------------------------------
\section{Conclusions}
This paper presents an investigation on end-to-end approaches for automatic speech assessment for Cantonese-speaking PWA. The CNN model trained with equal-length utterances using Log-Mel filterbank features outperforms the conventional two-step assessment method. The experimental results confirm the effectiveness of using CNN model for automatic extraction of pathological-related features. With the CAM visualization technique, it is observed that the CNN-learned features show similar physical meaning to human-designed features. It suggests that applying end-to-end approach is able to improve the efficiency of developing assessment system and save significant amount of manual work.
\begin{acknowledgements}
% This research is partially supported by the GRF project grants
% (Ref: CUHK14204014 and CUHK14227216) from the Hong
% Kong Research Grants Council, the Major Program of National
% Social Science Fund of China (Ref: 13\&ZD189), and the
% CUHK Shenzhen Research Institute.
This research is partially supported by a direct grant from Research Committee of the Chinese University of Hong Kong (CUHK) and a GRF project grant (Ref: CUHK14227216) from the Hong Kong Research Grants Council.
\end{acknowledgements}

\bibliographystyle{ieeetr}
\bibliography{reference2.bib}    % name your BibTeX data base

% % FOR BIBLATEX
% \printbibliography

% % Non-BibTeX users please use
% \begin{thebibliography}{}
% %
% % and use \bibitem to create references. Consult the Instructions
% % for authors for reference list style.
% % Format for Journal Reference
% Author, Article title, Journal, Volume, page numbers (year)
% Author, Book title, page numbers. Publisher, place (year)
% \end{thebibliography}

\end{document}